\documentclass[a4paper]{jpconf}
\usepackage{graphicx}
\begin{document}
\title{The M2-brane over the twisted torus with punctures}

\author{M.P Garcia del Moral$^{1,a}$, P Le\'on$^{1,b}$ and A Restuccia$^{1,c}$}
\address{$^{1}$Departamento de F\'isica, Universidad de Antofagasta, Aptdo 02800, Chile.}
\ead{$^{a}$maria.garcialdelmoral@uantof.cl, $^{b}$pablo.leon@ua.cl  and $^{c}$alvaro.restuccia@uantof.cl}

\begin{abstract}
We present the formulation of the bosonic Hamiltonian M2-brane compactified on a twice punctured torus following the procedure proposed in \cite{mpgm14}. In this work we analyse two possible metric choice, different from the one used in \cite{mpgm14}, over the target space and study some of the properties of the corresponding Hamiltonian.
\end{abstract}

\section{Introduction}

In a recent work \cite{mpgm14} we showed a formulation of the supermembrane theory in a background characterized by $M_9\times LCD$, where $LCD$ represents the one loop closed string interaction diagram with one incoming/outgoing string. This particular surface is conformally equivalent, through the Mandelstam map, to the twice punctured torus and it has only one compact direction, which is non trivial. Thus the formulation of the M2 brane on $LCD$ has ten non compact direction.  The non trivial topology of the target space induce non zero massive terms for all the dynamical fields of the theory as well a cosmological term on the Hamiltonian of the theory. The existence of massive terms ensures that if there exist a matrix regularization of the Hamiltonian it must have a discrete supersymmetric spectrum. Besides, the relation between the mapping class group of the twice punctured torus with the $(1,1) - Knots$ \cite{Cattabriga}, allows to classify the monodromies over the twice punctured torus into two subgroups. One of then related with the non trivial $(1,1)-knots$ and the other with trivial knots. The  trivial knots are related with the monodromies defined over the torus without punctures.

 In \cite{mpgm13} the authors show that the M2-brane on a torus bundle with parabolic monodromy can be decompactified to ten non compact dimensions such that it corresponds to a M2-brane on a twice punctured torus bundle. The authors require a restriction to parabolic monodromies to define the tenth decompactified direction. Thus the formulation of the M2-brane on a LCD can also be interpreted as the decompactification limit to ten non compact dimensions of the M2-brane on a torus bundle with fluxes \cite{mpgm6,mpgm10} and restricting to parabolic monodromy. This goes in concordance with Hull's conjecture \cite{hull8}, in which the M-theory origin of the massive Romans supergravity could be obtained from the T-dual of the nontrivial decompactification of the M-theory on a torus bundle with parabolic monodromy. Then, the formulation of the M2-brane on  $M_9\times LCD$ shown in \cite{mpgm14} could represent a concrete realization of this conjecture.
 
 The metric over the $LCD$ in the formulation presented in \cite{mpgm14} was chosen in such a way that the Hamiltonian of theory is well defined. In this work we will explore two different definitions of the LCD metric. One of then correspond to the usual metric defined over Riemann surfaces with punctures in the computation of string scattering amplitudes (see for example \cite{SONODA}). The second one appears in the analysis of magnetic monopoles over non trivial Riemann surfaces \cite{Restuccia2}. In section 2 we will present a brief discussion about the relation of the Riemann surfaces with punctures and the L.C string interaction diagrams. The section 3 will be focus on the formulation of the M2-brane on $LCD$ with the to choice for the target metric. Finally in section 4 we will present a discussion of the results.

\section{Riemann Surfaces with punctures and the Light Cone Diagrams}
 
 In this section we will make a short review about some of the quantities defined in Riemann surfaces with punctures and the equivalence with a Light Cone Diagrams for the closed strings interactions. 
 
 Let  $\Sigma_{g,N}$ be a Riemann surface of genus $g$ with $N$ punctures at  $Z_r$, with $r=1,..,N$ in a complex coordinate system $z$  defined over $\Sigma_{g,N}$. A known result from \cite{Mandelstam,Mandelstam2},(see also \cite{Giddings2}) is that the Riemann Surfaces  with punctures are conformally equivalent to the closed string interaction Light Cone Diagrams. The conformal map that relates the two surfaces, better known as Mandelstan's map, is given as 
 
 \begin{equation}
F(z)=\sum_r \alpha_r\left[ln E(z,Z_r)-2\pi i \int_{z_0}^{z} w \frac{1}{Im\Omega} Im\int_{z_0}^{Z_r} w \right] \quad \mbox{with} \quad \sum_r \alpha_r =0,
\end{equation} 
where $z_0$ is an arbitrary point on the surface (excluding the punctures) and $w=(w)_j$ with $j=1...g$ is the dual of the homology basis of holomorphic one forms, on the compact without punctures Riemann surface, that satisfy

\begin{equation}
\int_{a_j} w_i = \delta_{ij}, \quad \int_{b_j}w_i = \Omega_{ij}.
\end{equation}  
 Here the $a_j$,$b_j$ are the homology cycles of the Riemann surface and $\Omega_{ij}=\Omega$ is the period matrix. Finally $E(z,z_r)$ is a prime form defined for an odd spin structure and expressed in terms of the theta functions with characteristic $[s]$ as
 
\begin{equation}
 E(z,z_r)= \frac{\Theta[s]\int_{z_r}^z w|\Omega}{h_s(z)h_s(z_r)}, \quad \mbox{with} \quad h_s(z)= \sqrt{\sum_j \frac{\partial \Theta[s]}{\partial \xi_j}(0|\Omega)w_j(z)}.
\end{equation}  

The one form $dF$ is an Abelian differential of the third kind, since it has poles at the punctures. On the other hand $dF$ has also zeros at $P_a$ with $a=0,..,2g-2+N$ . The order of the poles and zeros depend of the interaction order in the L.C diagrams.  Now, $dF$ can be used to define an almost flat  metric over $\Sigma$ represented by the line element

\begin{equation}
    ds^2 = |\partial_zF|^2|dz|^2.
    \label{sm}
\end{equation}
This definition will break down at the punctures and zeros of the derivative of $dF$. 

Now, the set of parameters necessary to characterize $\Sigma_{g,N}$ are the Teichm\"uller parameter $\tau$ and the positions of the punctures $Z_r$. On the other hand, the set of parameters that describe the L.C diagrams are the external momenta $\alpha_r$, the internal momenta $\beta_j$, the interactions times $T_u$, with $(u=1,..,2g-3+N)$, and the twist angles $\theta_v$ with $(v=1,..,3g+N-3)$. However, not all these parameter are independent due to the momenta conservation condition.  The non trivial integrals of the one form $dF$ are given by

\begin{equation}
    \int_{a_j}dF = 2i\pi \beta_j, \quad \int_{C_r}dF = 2i\pi \alpha_r,
\end{equation}
where $C_r$ are curves around the punctures. The integral around the homology basis elements $b_j$ are more complicated to write in general due to the position of the punctures and the twist angles. Using the Mandelstam map it also possible to define a globally well behave coordinate, which in the string interaction contexts correspond to the light cone time, as

\begin{equation}
    T(z)\equiv \int^{z}_{z_0}Re(F)dz.
\end{equation}

Another important result about the L.C diagrams $\mathcal{X}$ with $n$ incoming strings and $m$ outgoing strings is that under small perturbations the diagram can be arranged in such way that all the interactions times are different. Then, to analyse  more complicated diagram $\mathcal{X}$, it is only required to study  simpler diagrams with only one incoming/ougoing string. Thus, for this kind of diagrams it is possible to establish a relation between the moduli of the surfaces given by 

\begin{equation}
\int_{b_j} dF - \sum_{i}^g \Omega_{ij}\int_{a_j}dF=2\pi i \int_{Z_2}^{Z_1}w_j.   
\end{equation}

Finally, to finish this section let us comment that, as we say before, the metric given  by (\ref{sm}) is very useful in the context of string interactions but for other theories, in which the conformal symmetry is not present, it could be more useful use another metric for the Riemann Surface. Another simple possibilities are 

\begin{eqnarray}
d\hat{s}^2 &=& dK^2 + dH^2, \label{m1} \\
d\bar{s}^2 &=& (K^2-1)dK^2 + (K^2-1)^{-1}dH^2, \quad K\equiv \tanh(G),  \label{m2} 
\end{eqnarray}
where $G\equiv Re(F)$ and $H\equiv Im(F)$. It can be proved that $G$ is a single valued function, however the one form $dG$ is harmonic since it has poles at the punctures. On the other hand $H$ is multivalued and $dH$ is harmonic. These two metrics have the sames issues of the metric (\ref{sm}), since both are defined in terms of the Mandelstam map.

\section{The M2-brane in a L.C diagram}

In this section we will follow the analysis presented in \cite{mpgm14} to write the Hamiltonian of the M2-brane in background characterized by $M_9 \times LCD$. Specifically, we will study this formulation particularized for the metrics  (\ref{sm}) and (\ref{m2}), since the M2-brane using (\ref{m1}) was already studied in detail in \cite{mpgm14}. In order to define the metric over the base manifold $\Sigma_{1,2}$ it is convenient to extract disk around the punctures and zeros of $dF$. Then the  determinant of the world volume metric $\sqrt{W}$ on $\Sigma_{1,2}$ can be defined as the pull-back of the volume form on $LCD$. 

\subsection{Using the metric $ds^2=|dF|^2$ over $LDC$}
\begin{figure}[h]
    \centering
\includegraphics[scale=0.3]{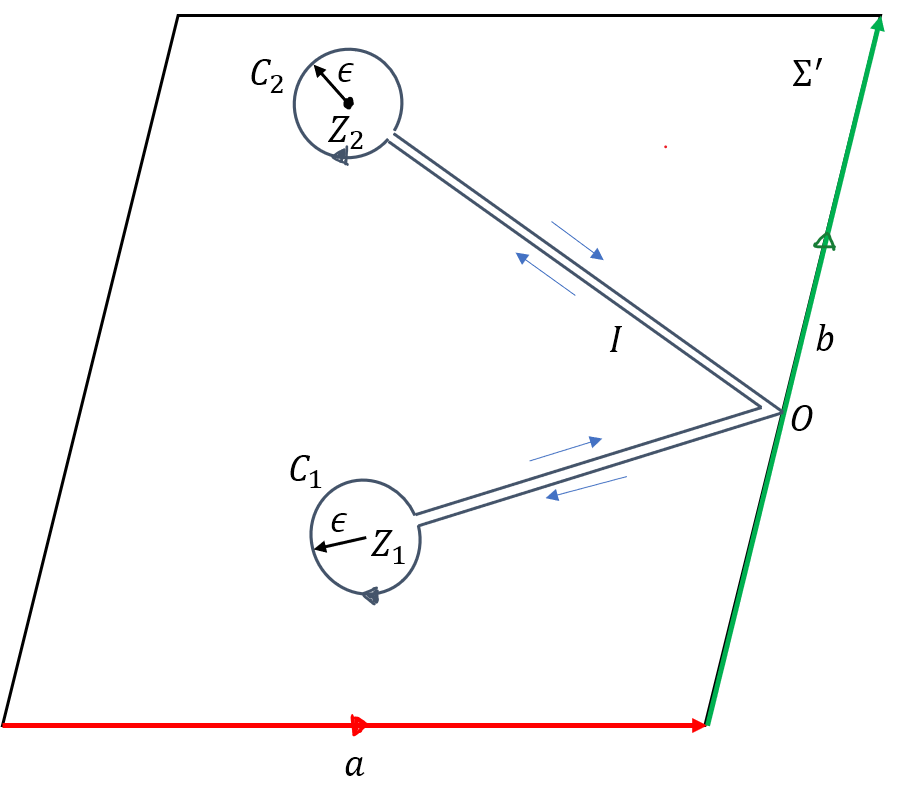}
    \caption{The  $\Sigma'$ region.}
    \label{fig:SigmaP}
\end{figure}

In this case the world volume metric over $\Sigma_{1,2}$ without the disks around the poles and zeros is given by 
\begin{eqnarray}
     \sqrt{W} &=& \epsilon^{ab}\partial_a G\partial_b H,
\end{eqnarray}
where $a,b=1,2$ are the spacial world volume coordinates. Now, we can define the M2-brane non trivial maps as
\begin{equation}
    dX^G=dG+dA^G, \quad dX^H=dH+dA^H,
\end{equation}
where the $dA^G$ and $dA^H$ are exact  one forms over $\Sigma_{1,2}$. Then, assuming that the dynamic fields $X^m, \ A^G$ and $A^H$ are well defined over the Riemann surfaces including the poles and zeros of $dF$, it can be show that 
\begin{eqnarray}
     \int_{\Sigma_{1,2}}\mathcal{H}=   \int_{\Sigma_{1,2}/S}\mathcal{H}, \quad \mbox{with} \quad S=(P_1,P_2). \label{zeros}
\end{eqnarray}
Then, as explained in \cite{mpgm14}, the zeros does not contribute to the Hamiltonian of the theory. In order to perform the integral taking into account the punctures, first we consider the surface $\Sigma_{1,2}$ as the fundamental region $\mathbf{\Sigma}_{1,2}\equiv \mathbf{R}^2 / \Gamma$, where $\Gamma$ is a discrete subgroup. Now, we can define the region $\Sigma'$  by cutting  $\mathbf{\Sigma}_{1,2}$ through a closed curve $c$ that goes around the punctures, with a radius $\epsilon$, and touch a point $O\in \partial \mathbf{\Sigma}_{1,2}$ (see figure \ref{fig:SigmaP}). This curve can be decomposed into the curves $C_1$,$C_2$ and $I$, where $I$ can be chosen as a curve $H=const$.  Thus, the bosonic Hamiltonian of the M2-brane can be written as \footnote{For simplicity we are taking the M2 brane tension $T_{M_2}=1$ in this work.}

\begin{eqnarray}\label{Hamiltonian massive}
     H &=& \lim_{\epsilon\rightarrow 0} \int_{\Sigma'} d^2\sigma \frac{1}{2}\sqrt{W} \bigg[ \left(\frac{P_m}{\sqrt{W}}\right)^2+\left(\frac{P_{G}}{\sqrt{W}}\right)^2+\left(\frac{P_{H}}{\sqrt{W}}\right)^2+\{X^m,X^{n}\}^2 \nonumber\\[0.3cm] &+&  \{G+A^{G},X^{n}\}^2
 +\{H+A^{H},X^{n}\}^2+ \{G+A^{G},H+A^{H}\}^2  \bigg].
\end{eqnarray}
From the last term of the Hamiltonian we obtain 
\begin{equation}
   \lim_{\epsilon\rightarrow 0} \int_{\Sigma'} \frac{1}{2}\sqrt{W}\{G,H\}^2 = -\lim_{\epsilon\rightarrow 0}  2\pi\ln(\epsilon) \rightarrow \infty, \label{log}
\end{equation}
which shows that the metric (\ref{sm}) leads to logarithm divergences in the Hamiltonian of the theory and for this reason this metric choice can not be used in this formulation of the M2-brane.

\subsection{Using the metric $ds^2=(K^2-1)^2dK^2 + (K^2-1)^{-2}dH^2$ over $LDC$}
Following the same steps of the previous case we define the metric over $\Sigma_{1,2}$ and the non trivial maps of the M2 brane as
\begin{eqnarray}
     \sqrt{W} &=& \epsilon^{ab}\partial_a K\partial_b H, \quad dX^K = dK+dA^K, \quad dX^H=dH+dA^H. 
\end{eqnarray}
Now, under the assumption that all dynamic fields are well defined over $\Sigma_{1,2}$, the Eq. (\ref{zeros}) holds for this case too. Then, using the definition of the momenta zero modes 
\begin{eqnarray}
    P_{K}=\sqrt{W}P_{0K}+\Pi_K, \quad P_{H} = \sqrt{W}P_{0H}+\Pi_H, \quad P_{m} = \sqrt{W}P_{0m}+\Pi_m,
\end{eqnarray}
the Hamiltonian can be expressed as
     \begin{eqnarray}\label{Hamiltonian massive 2}
     \small
     H &=& 2\pi \alpha + 2\pi \alpha \left(P_{0m}^2+\frac{2}{3} P^2_{0K}\right) + P_{0K} \lim_{\epsilon\rightarrow 0} \int_{\Sigma'} d^2\sigma \sqrt{W}\Pi_K (1-K^2)   \nonumber \\ &+& \lim_{\epsilon\rightarrow 0} \int_{\Sigma'} d^2\sigma \frac{1}{2}\sqrt{W} \bigg[ \left(\frac{\Pi_m}{\sqrt{W}}\right)^2+\left(\frac{\Pi_{K}}{\sqrt{W}}\right)^2\left(\frac{1}{1-K^2}\right)+\left(\frac{\Pi_{H}}{\sqrt{W}}\right)^2\left(1-K^2 \right)+\{X^m,X^{n}\}^2 \nonumber\\[0.3cm] &+&  \{K+A^{K},X^{n}\}^2\left(1-K^2\right)
 +\{H+A^{H},X^{n}\}^2\left(\frac{1}{1-K^2}\right)+ \{K+A^{K},H+A^{H}\}^2  \bigg],
\end{eqnarray}
where $P_{0H}$ have to be fixed to zero in order to avoid a logarithmic divergence of the form (\ref{log}).  Now, following the same procedure presented in \cite{mpgm14}, we can fix the gauge

\begin{eqnarray}
\{K,A^K\} + \{H,A^H\} = 0, 
\end{eqnarray}
and then, since it is identically zero, can be squared and added to the bosonic potential. The quadratic terms of the bosonic potential can be written as
\begin{eqnarray}
\small
    \mathcal{V} & = & \left(\frac{1}{1-K^2}\right)(\{K,X^m\}^2+\left(1-K^2 \right)\{H,X^m\}^2+  \{K,A^K\}^2+\{H,A^K\}^2   \nonumber \\ &+&\{K,A^H\}^2 + \{H,A^H\}^2.
\end{eqnarray}
Following the same steps of \cite{mpgm14}, it can be probed that 
\begin{eqnarray}
\left(\frac{1}{1-K^2}\right)\{K, X^m\}^2+\left(1-K^2 \right)\{H,X^{m}\}^2 &\ne& 0,  \label{mass1}\\
\{K,A^K\}^2+\{H,A^K\}^2 &\ne&  0, \label{mass2} \\ 
\{K,A^H\}^2+\{H,A^H\}^2 &\ne&  0, \label{mass3}
\end{eqnarray}
in $\Sigma_{1,2}/S$. Thus the bosonic Hamiltonian take the form

\begin{eqnarray}
\small
    H &=& 2\pi \alpha+ 2\pi \alpha \left(P_{0m}^2+\frac{2}{3} P^2_{0K}\right) + P_{0K} \lim_{\epsilon\rightarrow 0} \int_{\Sigma'}d^2\sigma \sqrt{W}\Pi_K (1-K^2) +2\pi \alpha\left[A^K(Z_2)-A^{K}(Z_1)\right] \nonumber \\
    & + & \lim_{\epsilon\rightarrow 0} \int_{\Sigma'}dK\wedge dH \frac{1}{2}\left[\left(\Pi_m\right)^2+\left(\Pi_K\right)^2\left(\frac{1}{1-K^2}\right)+\left(\Pi_H\right)^2\left(1-K^2 \right) + \frac{1}{2}\{X^m,X^n\}^2 \right. \nonumber \\&+&  \left(\frac{1}{1-K^2}\right) (\partial_H X^m +\{A^K,X^m\})^2+\left(1-K^2 \right)(\partial_KX^m-\{A^H,X^m\})^2 \nonumber \\ &+&  (\partial_HA^H+\{A^K,A^H\})^2+(\partial_KA^K+\{A^K,A^H\})^2 \nonumber \\
    & + & (\partial_KA^H)^2 + (\partial_HA^K)^2-\{A^K,A^H\}^2 +\{H+A^{H},X^{n}\}^2 \Bigg], \label{MM}
    \end{eqnarray}
subject to the constraints 
\begin{equation}
    d\chi = \int_a \chi = \int_b \chi = \int_{C_1} \chi = 0, \quad \mbox{with} \quad  \chi \equiv \left(\frac{P_K}{\sqrt{W}}\right)dX^K+\left(\frac{P_H}{\sqrt{W}}\right)dX^H+\left(\frac{P_m}{\sqrt{W}}\right)dX^m.
\end{equation}
This Hamiltonian has some of the properties of the Hamiltonian presented in \cite{mpgm14} with the metric (\ref{m1}) over $LCD$. In particular the existence of non vanishing  quadratic terms for all the dynamical field, the cosmological term associated with the width of the $LCD$, and a new global constraint which appears as a consequence of the punctures. However, in this case it is necessary fix $P_0H=0$ in order to avoid a singularity and there is a non trivial coupling of the zero and non zero modes in the third term of (\ref{MM}). Due to the factors of $(1-K^2)^{-1}$ in some of terms in the Hamiltonian it is necessary to make further assumptions about the dynamic fields in order to ensure that the Hamiltonian is finite. The supersymmetric extension of this Hamiltonian requires a deeper study since the metric over $LCD$ is not flat. 
\section{Conclusions}
We have analysed the formulation of the M2-brane theory on $M_9\times LCD$ with to different metric over the $LCD$. In first place we use the metric given by Eq (\ref{sm}), which appear in the computation of scattering amplitudes in string theory. In the case of the M2-brane, this metric leads to logarithmic divergences in the Hamiltonian that can not be avoided. For the second metric, we use the one that appears in the analysis of magnetic monopoles over compact Riemann surfaces. In this case we follow the procedure presented in \cite{mpgm14} and obtain a new bosonic Hamiltonian with a cosmological term and mass terms for each dynamic field. Now, in contrast with the Hamiltonian, obtained in \cite{mpgm14}, for the one presented here it is necessary to impose $P_{0H}=0$ and assume that the dynamical fields are such that the terms in the Hamiltonian with $(1-K^2)^{-1}$ are well defined to avoid singularities. Besides, there is also a non trivial coupling between the zero and non zero modes. Finally, the supersymmetric extension of the Hamiltonian in Eq. (\ref{MM}) it not clear since the curvature of the metric (\ref{m2}) is not zero. In the same way, the bosonic potential in (\ref{MM}), due to the mass terms, does not posses valleys. However, a more detail analysis is required to establish if the Hamiltonian satisfied the sufficiency conditions \cite{mpgm11} to have a discrete spectrum.      

\section{Acknowledgements}
P.L. is supported by the Projects ANT1756 and ANT1956 of the Universidad de Antofagasta. P.L want to  thanks to CONICYT PFCHA/DOCTORADO BECAS CHILE/2019-21190517. The authors M.P.G.M. and P. L also thank to Semillero funding project SEM18-02 from U. Antofagasta, and to the international ICTP Network  NT08 for kind support.


\section*{References}
\begin{thebibliography}{10}

\bibitem{mpgm14}
M.P.~Garcia Garcia~del Moral, P.~Leon, and A.~Restuccia.
\newblock {The Massive Supermembrane on a Knot}.
\newblock 1 2021.

\bibitem{Cattabriga}
Alessia Cattabriga and Michele Mulazzani.
\newblock (1, 1)-knots via the mapping class group of the twice punctured
  torus.
\newblock {\em Advances in Geometry}, 4(2):263 -- 277, 29 Mar. 2004.

\bibitem{mpgm13}
Maria~Pilar Garcia~del Moral and Alvaro Restuccia.
\newblock {10D Massive Type IIA Supergravities as the uplift of Parabolic
  M2-brane Torus bundles}.
\newblock {\em Fortsch. Phys.}, 64:398--402, 2016.

\bibitem{mpgm6}
M.~P. Garcia Del~Moral, C.~Las~Heras, P.~Leon, J.~M. Pena, and A.~Restuccia.
\newblock {M2-branes on a constant flux background}.
\newblock {\em Phys. Lett.}, B797:134924, 2019.

\bibitem{mpgm10}
M.P. Garcia~del Moral, C.~Las~Heras, P.~Leon, J.M. Pena, and A.~Restuccia.
\newblock {Fluxes, twisted tori, monodromy and $U(1)$ supermembranes}.
\newblock {\em JHEP}, 09:097, 2020.

\bibitem{hull8}
C.~M. Hull.
\newblock {Massive string theories from M theory and F theory}.
\newblock {\em JHEP}, 11:027, 1998.

\bibitem{SONODA}
Hidenori Sonoda.
\newblock Functional determinants on punctured riemann surfaces and their
  application to string theory.
\newblock {\em Nuclear Physics B}, 294:157 -- 192, 1987.

\bibitem{Restuccia2}
I.~Martin and A.~Restuccia.
\newblock {Magnetic monopoles over topologically nontrivial Riemann surfaces}.
\newblock {\em Lett. Math. Phys.}, 39:379--391, 1997.

\bibitem{Mandelstam}
S.~Mandelstam.
\newblock Interacting-string picture of dual-resonance models.
\newblock {\em Nuclear Physics B}, 64:205 -- 235, 1973.

\bibitem{Mandelstam2}
S.~Mandelstam.
\newblock {Dual - Resonance Models}.
\newblock {\em Phys. Rept.}, 13:259, 1974.

\bibitem{Giddings2}
Steven~B. Giddings and Scott~A. Wolpert.
\newblock A triangulation of moduli space from light-cone string theory.
\newblock {\em Comm. Math. Phys.}, 109(2):177--190, 1987.

\bibitem{mpgm11}
L.~Boulton, M.P. Garcia~del Moral, and A.~Restuccia.
\newblock {Discreteness of the spectrum of the compactified D = 11
  supermembrane with nontrivial winding}.
\newblock {\em Nucl. Phys. B}, 671:343--358, 2003.

\end{thebibliography}
\end{document}